\documentclass[aps,pre,reprint,nofootinbib,amsmath,amssymb]{revtex4-2}

\usepackage{graphicx}\graphicspath{{figures}}
\usepackage{bm}
\usepackage{dcolumn}
\usepackage{esint}
\usepackage{subcaption}
\usepackage{braket}
\usepackage{lettrine}
\usepackage[utf8]{inputenc}
\usepackage[T1]{fontenc}
\usepackage{mathptmx}
\usepackage{appendix}
\usepackage{etoolbox}
\usepackage{lineno}

\begin{document}

\title{Theory of electron and ion holes as vortices in the phase-space of collision-less plasmas}
	\author{Allen Lobo}
	\email{allen.e.lobo@outlook.com}
	\affiliation{Sikkim Manipal Institute of Technology, Sikkim Manipal University, Sikkim - 737136, India.}
	\author{Vinod Kumar Sayal}
 \email{vksayal@hotmail.com}
	\affiliation{Sikkim Manipal Institute of Technology, Sikkim Manipal University, Sikkim - 737136, India.}
	\date{\today}
 
	\begin{abstract}
    This article studies the vortical nature and structure of phase-space holes -- nonlinear B.G.K. trapping modes found in the phase-space collision-free plasmas. A fluid-like outlook of the particles' phase-space is explored, which makes it convenient to analytically identify electron and ion holes as vortices -- similar to that of ordinary two-dimensional fluids. A fluid velocity is defined for the phase-space of the electrons and ions, continuity and momentum equations describing the flow of the phase-space fluid representing the particle system are then developed. Pressure formation and associated diffusion in phase-space of such systems is introduced and a vorticity field of the phase-space is then defined. Using these equations, electron holes and ion holes are analytically identified as vortices in the phase-space of the plasma. A relation between Schamel's trapping parameter ($\beta$), hole speed ($M$), hole phase-space depth ($-\Gamma$) and hole potential amplitude ($\chi_0$) is derived. The approach introduces a new technique to study the phase-space holes of collision-less plasmas, allowing fluid-vortex-like treatment to these kinetic structures. Phase-space distribution functions for electron hole regions can then be analytically derived from this model, reproducing the schamel-df equations and thus acting as a precursor to the pseudo-potential approach, avoiding the need to assume a solution to the phase-space density.
	\end{abstract}
	\maketitle
    \section{Introduction}
	\lettrine{I}{n} the study of collision-less electrostatic plasmas, the problem of electron and on holes stand amongst the most complex topics one has to deal with. Electron and ion holes, collectively also referred to as phase-space holes, are well-identifiable kinetic structures formed in the phase-space of collision-free electrostatic plasmas, governed by the Vlasov-Poisson equation system. With the core (inner) phase-space density much lower than the boundary region, these nonlinear structures represent B.G.K. modes \cite{Bernstein1957} with solitary potential waveform resembling a Gaussian wave, as seen in laboratory experiments and simulation studies \cite{morse1969one, turikov1978computer, Lynov1979, Lynov1980, Guio2003, Aravindakshan2021StructuralPlasma,  Eliasson2006}. They are formed as a result of nonlinear particle-trapping mechanisms \cite{Schamel2023PatternEquilibria} and travel with phase-space speeds (M) comparable to the particle thermal speed \cite{Iizuka1987,Lynov1979,Saeki1979}. The study of these phase-space holes is conducted using the differential or integral B.G.K. approaches \cite{Bernstein1957}, the former employing an assumed distribution function for the trapped and free particle regions in phase-space \cite{Schamel1979} and the latter an assumed potential function resembling the typical inverted bell shape \cite{Turikov1984, Chen2002Bernstein-Greene-KruskalPlasmas, Aravindakshan2021StructuralPlasma}.\\
    \indent Interestingly, electron and ion phase-space holes have been ubiquitously referred to as phase-space vortices since as early as their first observations via computer simulations. \textcite{morse1969one} reported the formation of electron holes from two-stream conditions and first stated them as phase-space vortices. The experimental and simulation study conducted by  \textcite{Saeki1979}, \textcite{Lynov1979} and later \textcite{Guio2003} also termed the electron holes formed in their experiment as phase-space vortices. Similar reference to the holes as phase-space vortices were presented in various analytical works by \textcite{schamel1986electron,Schamel1998} and other authors \cite{Turikov1984, Trivedi2017,Chen2005,Aravindakshan2021StructuralPlasma}. However, an analytical proof of the vortical nature of the electron and ion phase-space holes has not been presented still, to the best of our knowledge.\\
    \indent There is another issue with the present theory of electron (and ion) phase-space holes. Schamel's pseudopotential method \cite{Schamel1979} utilizes the schamel-df equations \cite{Schamel1979, schamel1986electron} to describe the structure of the electron hole both for free and trapped particle regions. These analytically assumed equations contain one arbitrary electron-trapping parameter $\beta$, which is described as the inverse of the trapped electron temperature \cite{Schamel1979, hutch2017}. To the best of our knowledge, this trapping parameter has not been described still as a measurable quantity.\\
    \indent In this article, the electron phase-space of collision-less plasmas is studied in analogy to a two-dimensional fluid. In section II, the $1D-1V$ phase-space of a many-particle system is premeditated as a two-dimensional fluid surface. Based on this, we develop continuity and momentum equations of the phase-space fluid which collectively represent the fluid-like behaviour of the phase-space. A vorticity field of the phase-space is then described. In section III, we analyse this vorticity field of the plasma particle phase-space and analytically identify electron and ion holes as vortices in the phase-space of the collision-less plasmas. Further, this vortical nature of the phase-space is verified using kinetic Vlasov simulation of electron hole formation in a Q-machine plasma. In section IV of this work, using the above-stated theory, we reproduce the schamel-df equations \cite{Schamel1979}. We also obtain a relation from which the electron trapping parameter $\beta$ can be directly measured in terms of the hole-depth in phase-space. The last section is then devoted to the conclusion of this work. As will be evident from the mathematics presented in this work, the fluid-analogue theory of the phase-space described herein encompasses a more thorough understanding of the phenomena of the phase-space holes.
    \section{Theory of phase-space vortices}
    The development of vortical structures from two-stream-like interactions and curling deformation is ubiquitous in ordinary fluid systems. Vortices formed in such systems present themselves with an abundant literature in their mathematical analyses. The correspondence between ordinary two-dimensional fluids and the $1D-1V$ is also well-known \cite{schamel1986electron, schamel2012cnoidal}, with the Vlasov equation of the collision-less plasma fluid being identical to the Liouville equation \cite{Louiville1838} of the general case. For a theoretical analysis of the phase-space vortices, the phase-space itself must be studied in analogy to a two-dimensional fluid, since this vortical nature is exhibited in the phase-space and not the one-dimensional configuration space, in which these kinetic structures are represented by the B.G.K. solitary potential waves. In this section, we explore this analogy between an ordinary two-dimensional fluid and a $1D-1V$ phase-space and derive a set of equations in phase-space which are analogous to the Euler equations of the fluid case. These equations can then be utilized for a fluid-like analysis of the phase-space. We also describe the presence of a statistical, kinetic pressure in the phase-space which develops due to the probability density gradient along position space. We then develop a vorticity field to describe the curling deformation of this phase-space fluid surface. \\
  \indent The position-velocity phase-space of a system can be described as a two-dimensional surface by multiplying a system-specific characteristic time to the velocity axis. We represent it as $\tau$ for our analysis. The phase-space then describes a co-ordinate set $(x-\tau v_x)$ to locate an infinitesimal volume $\tau dx dv_x$. The particle density of this phase-space is the number of particles present in this volume at any given instant -- 
  \begin{equation}\label{psdensity}
     \frac{\partial^2 N}{\tau\partial x \partial v_x}=\frac{1}{\tau}f(x,v_x,t) =\eta.
  \end{equation}
  Here, $N$ is the total number of particles in the system. The phase-space gradient operator can similarly be defined as the derivative along each axis --
  \begin{equation}\label{nablameaning}
      \nabla = \frac{\partial}{\partial x}\hat x + \frac{\partial}{\tau\partial v_x}\hat v_x.
  \end{equation}
  One can now describe the associated position $\bm{r}$ and velocity field $\mathcal{V}$ of a phase-space fluid element in phase-space --
  
  \begin{subequations}\label{vnjfields}
      \begin{align}
      \bm{r} = x\hat x + \tau v_x \hat v_x,\\
          \mathcal{V} = \frac{d\bm{r}}{dt} = \dot x \hat x + \tau \dot v_x \hat v_x.\\
          \intertext{The flux density of the flow in the phase-space $\bm{J}$ will then be the multiple of the particle density and velocity field of the phase-space --}
          \bm{J} = \eta \mathcal{V} = \frac{1}{\tau}f\dot x \hat x + f\dot v_x \hat v_x.
      \end{align}
  \end{subequations}
  
\begin{figure*}[ht]\centering
		\begin{subfigure}{2.3in}	\includegraphics[width=1.0\linewidth]{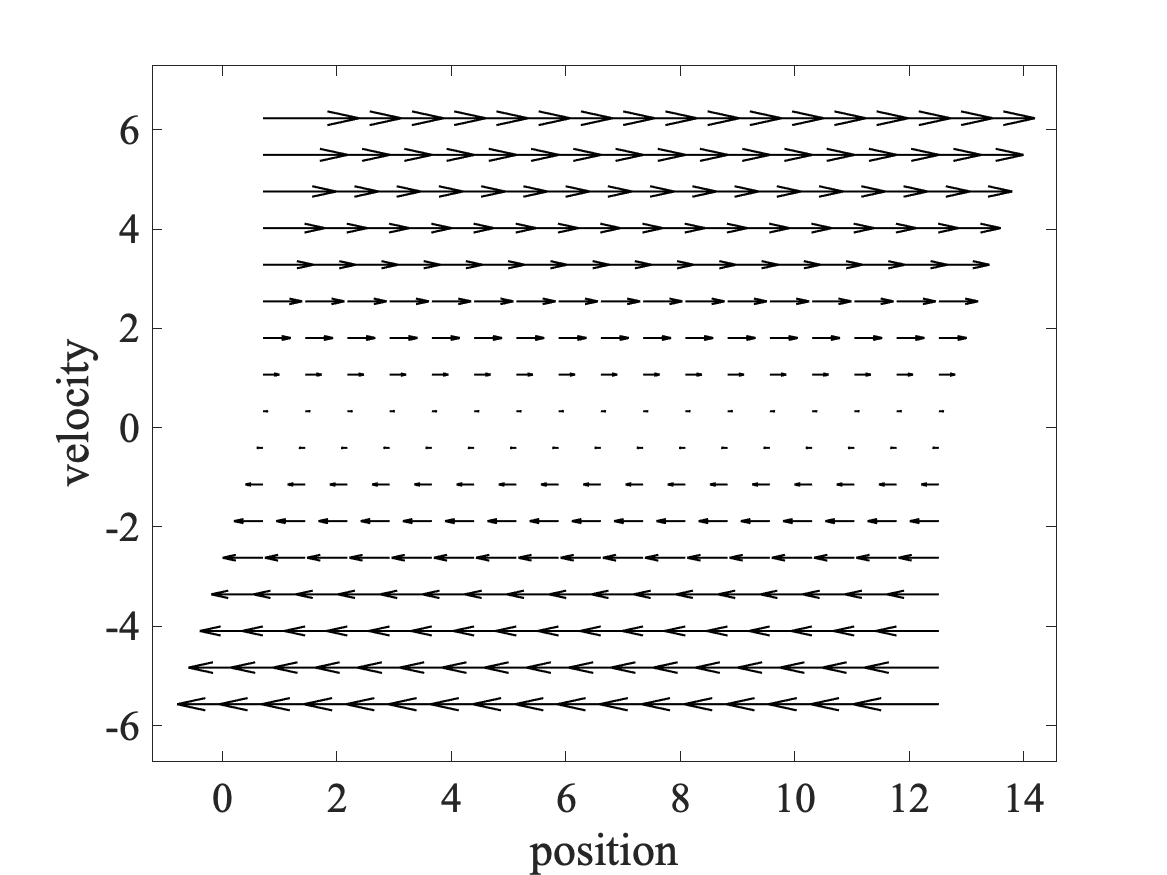}
			\caption{}
		\end{subfigure}
	\begin{subfigure}{2.3in}
		\includegraphics[width=1.0\linewidth]{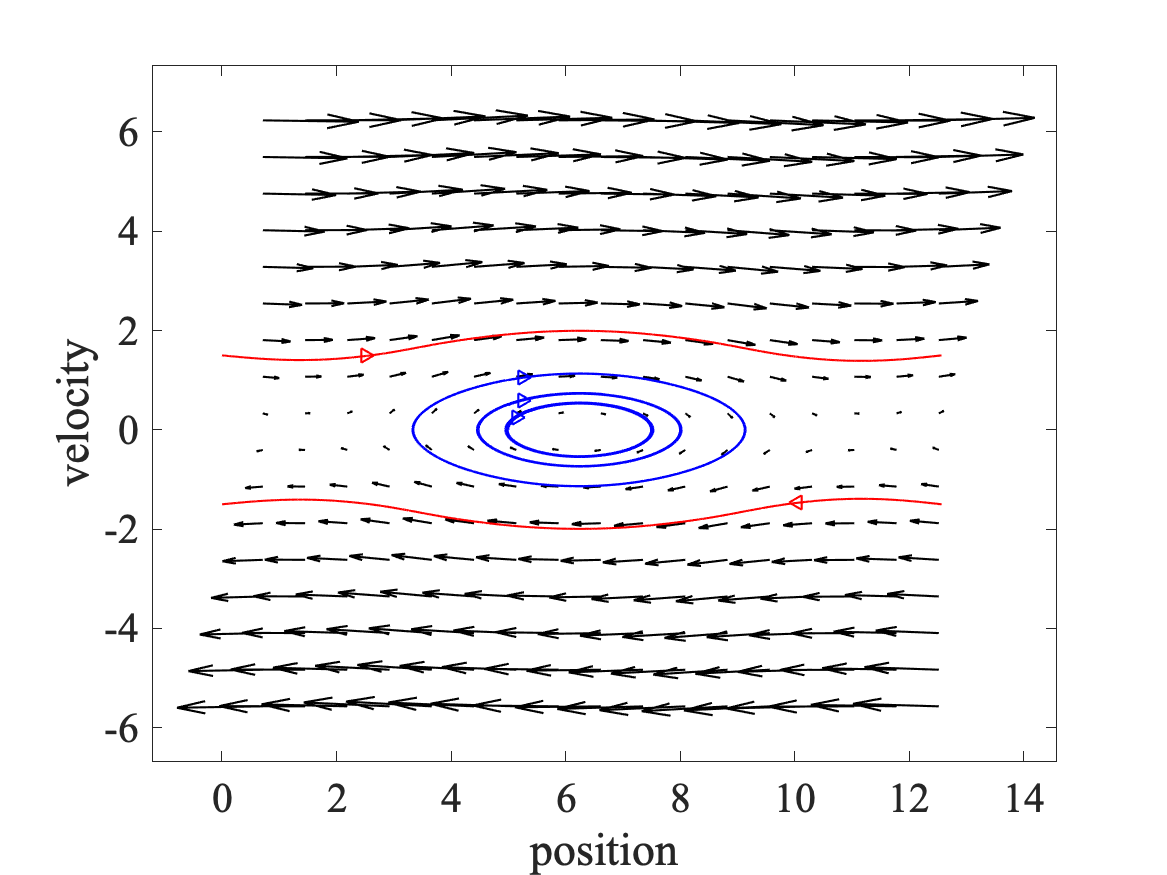}
		\caption{}
	\end{subfigure}
 \begin{subfigure}{2.3in}
		\includegraphics[width=1.0\linewidth]{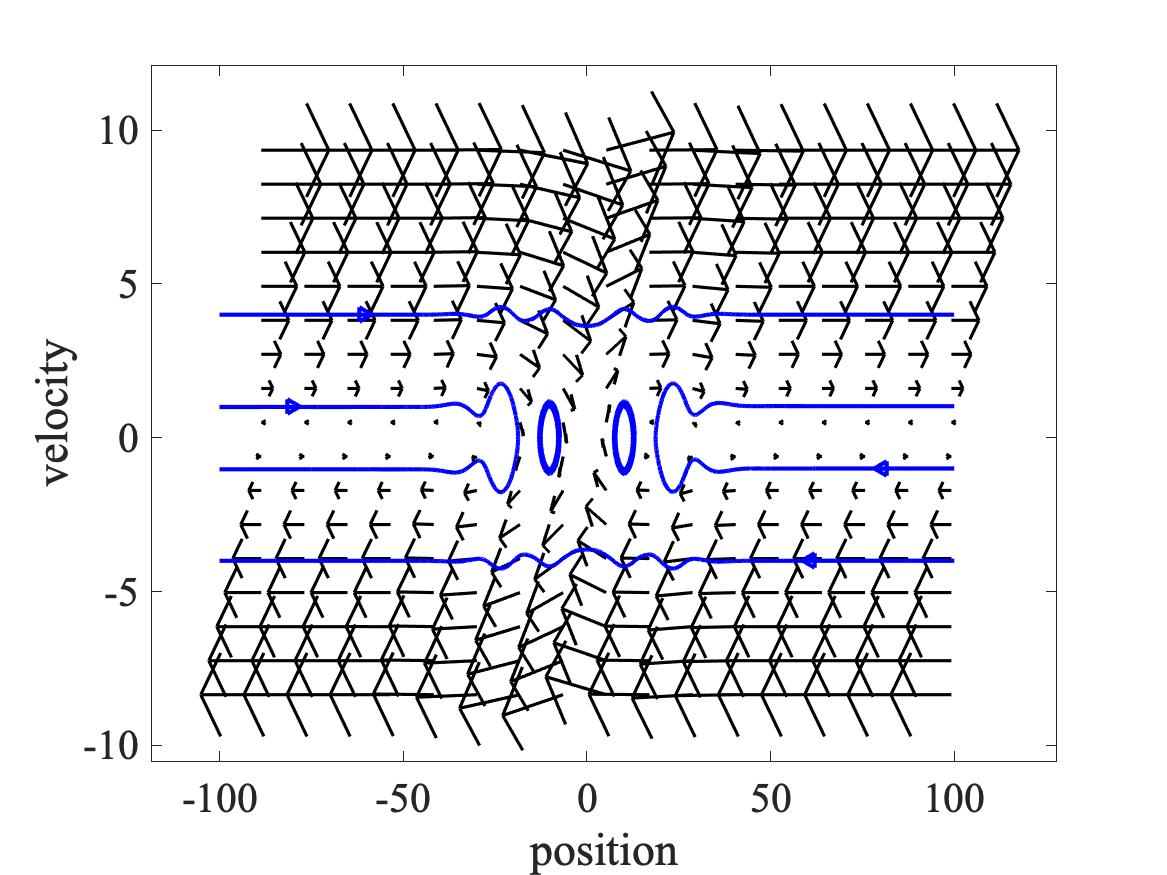}
		\caption{}
	\end{subfigure}
\caption{Velocity field $\mathcal{V}$ plots of particle phase-space of an ideal system (a) with Maxwellian (thermal equilibrium) distribution and (b) in the presence of an negative bell-like attractive potential with streamlines showing vortical motion of the phase-space fluid resembling the phase-space trajectories of trapped (blue) and free (red) phase-space fluid elements, (c) Unsteady flow in presence of hyperbolic secant potential waves, flow represented by various streamlines. }
\label{Figure1}
	\end{figure*}

  Having defined the necessary fluid-like dynamical variables of the particle phase-space, flow-like dynamics of the phase-space can be portrayed. Self-consistency of this flow-like behaviour of the phase-space (see fig. \ref{Figure1}) is further defined upon describing a streaming function $\mathcal{H}$\footnote{Please refer to Appendix A, equations (A.1-A.3d) for the associated mathematics.} which can be described as the Hamiltonian $H$ times the constant $(\tau/m)$. using the  the kinetic-Euler equations which we develop in the next part of this section.
  
  \subsection{The continuity equation of phase-space and in-compressible flow of collision-less systems}
  The velocity and flux density fields of the particle flow in phase-space, described in equations (\ref{vnjfields}a-c) can now be used to understand the nature of the flow of this phase-space fluid. We begin by studying the time-rate of change of the particle density in the phase-space --
  \begin{subequations}\label{continuityeqn}\allowdisplaybreaks
      \begin{align}
          &\frac{d\eta}{dt} = \frac{1}{\tau}\left(\frac{\partial f}{\partial t}+\frac{\partial f}{\partial x}\frac{dx}{dt}+\frac{\partial f}{\partial v_x}\frac{dv_x}{dt}\right).\\
          \intertext{Also,}
          &\nabla\cdot\mathcal{V} = \frac{\partial \dot x}{\partial x} + \frac{\partial \dot v_x}{\partial v_x} = \frac{\partial}{\partial x}\frac{\partial H}{\partial v_x}-\frac{\partial}{\partial v_x}\frac{\partial H}{\partial x} = 0.\\
          \begin{split}
               \nabla\cdot\bm{J} = \eta \nabla.\mathcal{V} + \mathcal{V}.\nabla \eta = \dot x \frac{\partial \eta}{\partial x} + \tau \dot v_x \frac{\partial \eta}{\tau\partial v_x}\\
               =\frac{1}{\tau}\dot x\frac{\partial f}{\partial x} + \frac{1}{\tau}\dot v_x\frac{\partial f}{\partial v_x}.
          \end{split}\\
          &\Rightarrow \frac{d\eta}{dt} = \frac{\partial \eta}{\partial t} + \nabla\cdot\bm{J} =\begin{cases}
              0&\text{collision-less system}\\
              \left(\frac{\delta \eta}{\delta t}\right)_{\text{coll.}}&\text{with collisions}
          \end{cases} 
      \end{align}
  \end{subequations}
 Equation (\ref{continuityeqn}d) describes the continuity equation of the phase-space flow in the cases of collision-less and collisional considerations of the system \cite{Gibbs1902, Vlasov1938}. The flow of the phase-space fluid representation of the one-dimensional system is in-compressible in a collision-free condition.\\
 \indent The dynamics of the phase-space fluid analogue of the system can be further explored by framing phase-space momentum equations governing its flow. We do this by determining the time-derivatives of the fluid's velocity field. For the same, we introduce a general force field term $\bm{F}(x)$ and a kinetic pressure $\bm{P}$ into the system. Wherein the force field $\bm{F}(x)$ represents the net interactions of the particles with any internal and external force fields influencing the system, the theory of a pressure in phase-space of the system is related to its stochastic relaxation when initially disturbed. We momentarily digress from the momentum equation derivation and explore this pressure present in the phase-space of the system.
 \subsection{Pressure in the phase-space of an initially disturbed system.}
 The theory of a pressure formation in phase-space develops from the affinity of many-particle systems to attain a state of statistical equilibria. In a system with an initial non-uniform spatial particle distribution, particle diffusion shifts this distribution towards uniformity. Similarly, the distribution function $f(x,v_x)$ and hence the phase-space fluid particle density $\eta$ tends towards a state of spatial uniformity over time, from an initially disturbed or non-uniform state. This diffusive flow of particles in the phase-space from regions of higher probability density towards regions of lower probability density must be accompanied by the formation of a pressure $\bm{P}$ which simply directs the flow of the particles towards a relaxed probability distribution $f(x,v_x,t)$, and a diffusion current $\bm{J}_{\text{diff.}} = -\mathcal{D}\nabla\eta$, analogous to diffusion in ordinary fluids, representing this flow, where $\mathcal{D}$ is the phase-space diffusion coefficient. Therefore, the kinetic pressure gradient $\nabla_x\bm{P}$ should be opposite and proportional to the phase-space density gradient --
 \begin{equation}\label{kp}
     \nabla_x\bm{P} \propto -\nabla_x\eta
 \end{equation}
 \begin{figure}[!h]
    \centering
    \includegraphics[width=0.8\linewidth]{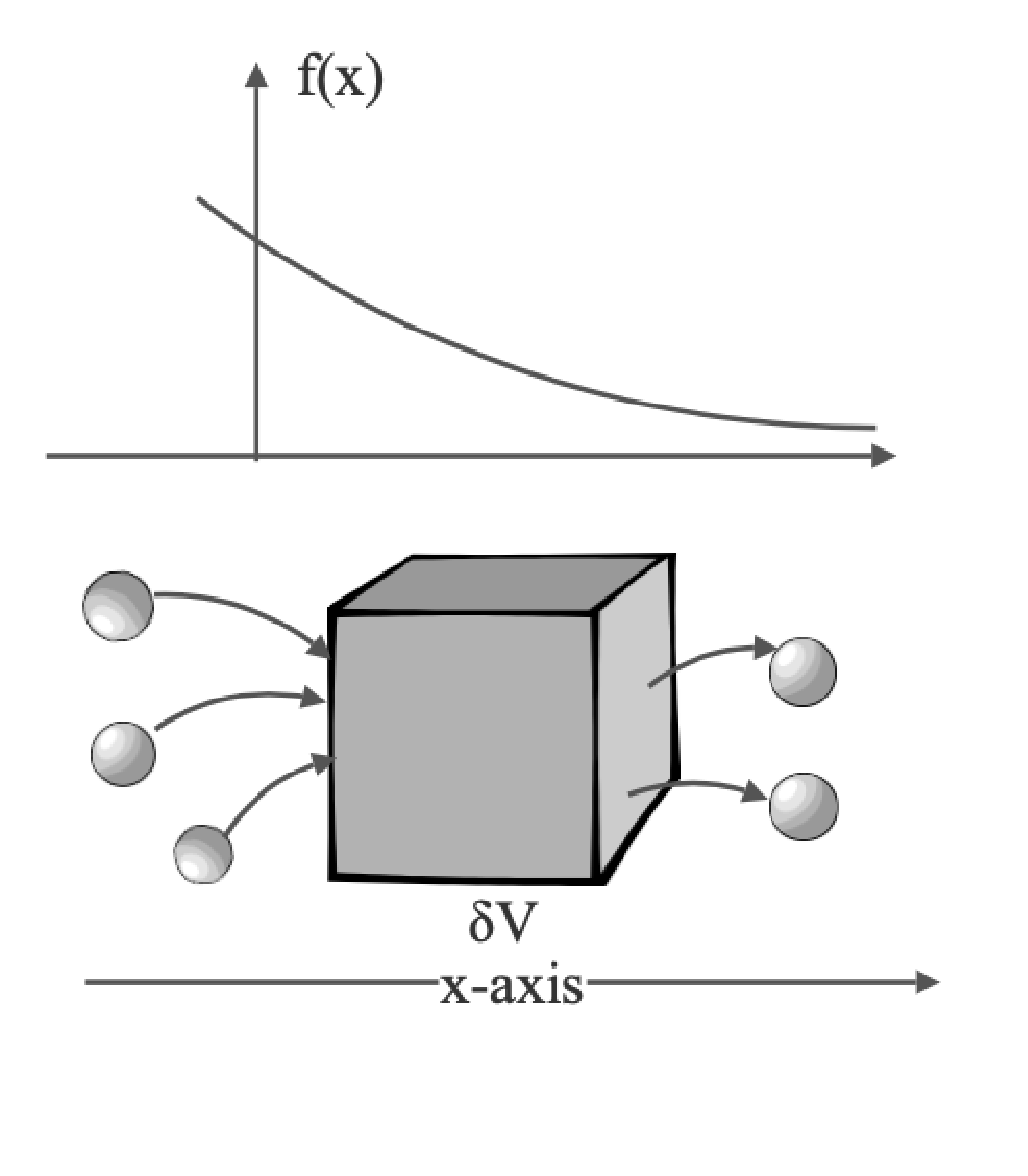}
    \caption{net flow of particles across a phase-space volume element $\delta V$, along the decreasing probability density $f(x,v_x)$.}
    \label{fig:kinpre}
\end{figure}
The phase-space pressure originates due to the net flux of particles across a phase-space volume element $\delta V = \tau \delta x\delta v_x$. Owing to the non-uniform spatial distribution along position axis, the flow current density $\bm{J}_x$ will be higher when directed into this volume $\delta V$ than directed out of it. Hence, the net flux of particles must be negative at any given instant. Assuming a classical equilibrium distribution of particles along the velocity axis, the energy of each particle becomes $K_BT$. The net energy flux across $\delta V$ is therefore $K_BT_i$ times the net particle flux across this volume, $-\delta n$. The net work $\delta W$ done during this flow by the phase-space pressure in this volume then becomes --
\begin{subequations}
    \begin{align}
        \delta W = \bm{P}\delta V = -\delta n\cdot K_BT_i\\
        \Rightarrow\bm{P} = -\frac{\delta n}{\delta V}K_BT_i=-K_BT_i\eta
    \end{align}
\end{subequations}
In this work, we assume a Maxwellian, isothermal form of this pressure. However, other forms may also exist in other systems, such as with non-uniform temperature distributions. The pressure gradient in the phase-space fluid analogue similarly becomes --
 \begin{equation}\label{kineticpressuregrad}
     \nabla_x\bm{P} =-K_BT_i\nabla_x\eta
 \end{equation}
This kinetic pressure gradient, which exists along the position axis in the phase-space and exists due to the uneven probability distribution or phase-space density along the position space, can be used to further construct the kinetic momentum equation along the position space, which we perform in the next part of this work.

 \subsection{Kinetic momentum equation along position in particle phase-space}
 The time-rate of change of the position component of the fluid velocity $\mathcal{V}$ of the phase-space fluid analogue can be written as --
 \begin{subequations}\allowdisplaybreaks\label{Eulereqns}
     \begin{align}
         &\frac{d\mathcal{V}_x}{dt}=\Ddot{x}\hat x.\intertext{Introducing advection terms,}
		&\ddot x(x,t) = \frac{\partial \dot x}{\partial t} + v_x\frac{\partial \dot x}{\partial x}.\\
         \intertext{For a system experiencing an internal force field $\bm{F}(x,t)$ and kinetic pressure $\bm{P}$,}
        &\Ddot{x} = \frac{\bm{F}(x,t)}{m}-\frac{\nabla_x \bm{P}}{m\eta}.\\
        \intertext{Inserting equations (\ref{Eulereqns}b) and (\ref{Eulereqns}c) into equation (\ref{Eulereqns}a), we get --}
       &\frac{\partial \dot x}{\partial t} + v_x\frac{\partial \dot x}{\partial x} = \frac{\bm{F}(x,t)}{m} -\frac{1}{m\eta}\frac{\partial \bm{P}}{\partial x} \\
       &\frac{\partial \dot v_x}{\partial t} +  v_x\frac{\partial \dot v_x}{\partial x}=\frac{\dot{\bm{F}}(x,t)}{m}. 
     \end{align}
 \end{subequations}
 The above equations (\ref{Eulereqns}d, e) describe the momentum equations for the flow of the phase-space fluid model of the system.
 For various particles, the equations can be generalised as --
 \begin{equation}\label{kineticeuler1}
     \begin{split}
         m_i\eta_i\left(\frac{\partial \dot x}{\partial t} + v_x\frac{\partial \dot x}{\partial x}\right) = \eta_i\bm{F_i}(x,t) -\nabla_x \bm{P}, \\
        m_i\eta_i\left(\frac{\partial \dot v_x}{\partial t} + v_x\frac{\partial \dot v_x}{\partial x}\right)=\eta_i\dot{\bm{F_i}}(x,t).
     \end{split}
 \end{equation}
 One can introduce more terms on the right side of the equations (\ref{kineticeuler1}) for other forms of interactions exhibited in phase-space. Equations (\ref{continuityeqn}d) and (\ref{kineticeuler1}) together describe the kinetic Euler equations for the fluid model of the particle phase-space. These equations are in capacity to collectively describe the complete dynamics of the system in its phase-space, in the same nature as the classical Euler equations describe the behaviour of an ordinary fluid. For the collision-less condition, equation (\ref{continuityeqn}d) can be set to zero. 
\subsection{Particle diffusion in phase-space}
  Particle diffusion in the phase-space fluid analogue can be observed as a consequence of the developed kinetic Euler equation by employing the phase-space pressure. For this, we study the kinetic momentum equation along position-space, with a time-invariant, quasi-static flow of the fluid in phase-space, therefore equating $\frac{\partial \dot x}{\partial t}$ to 0. Using $\dot x = v_x$, the particle velocity, the equation becomes --
 \begin{subequations}\label{diff1}\allowdisplaybreaks
 \begin{align}
     m_i\eta_i v_x\frac{\partial v_x}{\partial x}= \eta_i\bm{F_i}(x) -K_BT_i\nabla_x \eta_i\\
     \bm{J}_{x_i}=\eta_i v_x= \frac{\eta_i \bm{F}_i(x)}{m_i\frac{\partial v_x}{\partial x}} -\frac{K_B T_i}{m_i\frac{\partial v_x}{\partial x}}\nabla_x \eta_i.
 \end{align}
 \end{subequations}
Upon substituting $\dot v_x$ with $\bm{F}/m_i$ in the continuity equation (\ref{continuityeqn}b) of the phase-space, we get --
 \begin{equation}\label{dvdxsubstitution}\allowdisplaybreaks
     \begin{split}
         \nabla.{\mathcal{V}} = \frac{\partial \dot x}{\partial x}+\frac{1}{m_i}\frac{\partial \bm{F}}{\partial v_x}=0 \Rightarrow
        \frac{\partial v_x}{\partial x} = -\frac{1}{m_i}\frac{\partial \bm{F}(x)}{\partial x}\frac{\partial x}{\partial v_x}\\
         \Rightarrow \frac{\partial v_x}{\partial x} = \sqrt{\frac{1}{m_i}(-\nabla_x \bm{F})}.
     \end{split}
 \end{equation}
 Including equation (\ref{dvdxsubstitution}) in equation (\ref{diff1}b), we get --
 \begin{subequations}\label{diff2}\allowdisplaybreaks
     \begin{align}
     \begin{split}
        &\bm{J}_{x_i}=\eta_i v_x= \frac{\eta_i \bm{F}_i(x)}{\sqrt{-m_i\nabla_x \bm{F}}} -\frac{K_B T_i}{\sqrt{-m_i\nabla_x \bm{F}}}\nabla_x \eta_i,\\
         &\text{or,}\quad\bm{J}_x = \mu\eta_i\bm{F}(x) - \mathcal{D}\nabla_x \eta_i
         \end{split}\\
         &\text{where,} \quad \mu =\sqrt{\frac{-1}{m_i\nabla_x\bm{F}(x)}}, \quad \mathcal{D} = K_BT_i\sqrt{\frac{-1}{m_i\nabla_x\bm{F}(x)}}\\
         &\text{and}\qquad K_BT_i\mu = \mathcal{D}.
     \end{align}
 \end{subequations}
 In the above equations, $\mu$ represents particle mobility in phase-space along position ($x$) axis. $\mathcal{D}$ represents the fluid diffusion coefficient in phase-space, which is related to the mobility $\mu$ using Einstein's classical diffusion relation \cite{Einstein1905UberTeilchen, Sutherland1905LXXV.Albumin} shown in equation (\ref{diff2}c). Both particle mobility and diffusion coefficient are facilitated by the presence of a negative force-field gradient in these regions. Further understanding of this phenomena is therefore possible by introducing field-specific equations which describe this gradient. As an example, upon introducing an electrostatic interaction, $\bm{F} = q\bm{E} $ where $\bm{E}(x)$ is the electric field of this interaction, the field gradient will then become the spatial charge density, since, $\epsilon_0\nabla_x\bm{E} = \rho(x)$, suggesting that in an electrostatic domain, such a diffusion occurs in the presence of a charge accumulation in space.
 \subsection{Vorticity field in phase-space}
 \indent We now proceed to describe the vorticity field $\bm{\xi}(x,\tau v_x)$ associated with the phase-space fluid analogue of a system, which is simply the curl of the velocity field\footnote{Also refer to Appendix (A.4a-b) for an alternate determination of the phase-space vorticity using the Laplacian of the streaming function.}.
 \begin{subequations}\label{vorticity}
 \begin{align}
     &\Bigl|\nabla \times \mathcal{V}\Bigr| = \frac{\partial \tau \dot v_x}{\partial x} - \frac{1}{\tau}\frac{\partial \dot x}{\partial v_x}=\frac{\tau}{m_i}\frac{\partial \bm{F}(x)}{\partial x} - \frac{1}{\tau}\frac{\partial v_x}{\partial v_x}.
     \intertext{Here, $\dot v_x$ has been substituted by $\bm{F}(x)/m_i$.}
     \begin{split}
     \therefore\quad&\bm{\xi}= \left(\frac{\partial\tau \dot v_x}{\partial x} - \frac{1}{\tau}\frac{\partial \dot x}{\partial v_x}\right)\hat n=\left(\frac{\tau}{m_i}\frac{\partial \bm{F}(x)}{\partial x} - \frac{1}{\tau}\right)\hat n,\\
     &\text{where,}\qquad\hat n = \hat x \times \hat v_x.
     \end{split}
     \end{align}
 \end{subequations}

 Here, $\hat n$ is a direction normal to the phase-space plane. It is observable that vorticity of the phase-space fluid, which varies only along the position-space, is influenced by the present force-field gradient, without which it remains a constant. It is hence in the nature of this phase-space fluid model of a system to exhibit some type of local deformation, sheering or rotational, in time. Presence of vorticities in either direction, regulated by the direction of the force-field gradient, can subsequently give rise to local turbulence which can then result in formation of various characteristic structures, such as vortices (refer to FIG. \ref{Figure1}c).\\
\indent The equations developed in this section explore the flow-like behaviour of the phase-space of a many-particle system, and can now be utilized to identify and analyse vortical structures in the phase-space plane, which is done in the next section.
\section{Identification of electron and ion holes as phase-space vortices}
With reference to some established techniques of vortex identification in two-dimensional fluids \cite{Lugt1979, Tian2018a}, we have developed an analogous analytical technique to identify vortices in the phase-space of collision-less electrostatic plasmas. Specifically, we analyse the phase-space vorticity fields and observe particle trajectories in the phase-space of the plasma, along-with the local concentrations of vorticities as well as locally phase-space rotating fluid.\\
\indent Upon setting particle charge $q= \delta e$ with $\delta = \pm1$ for (singly ionised, positive or negative) ions or electrons, $\bm{F}=qE$  with $E(x,t)$ as electric field, $\dot x = v_x$ and $\dot v_x = \frac{\delta e}{m}E(x,t)$ in equation (\ref{vorticity}), the vorticity associated with a plasma particle (ion or electron) phase-space fluid flow becomes --
\begin{equation}\label{vorticityrho}\allowdisplaybreaks
	\begin{split}
	\bm{\xi}=&\Bigl(\tau\frac{\partial \dot v_x}{\partial x} - \frac{1}{\tau}\Bigr)\hat n = \Bigl(\frac{\delta e\tau}{m}\frac{\partial E(x,t)}{\partial x} - \frac{1}{\tau}\Bigr)\hat n\\
	=& \Bigl(\frac{\delta e\tau}{\epsilon_0m}\rho(x,t) - \frac{1}{\tau}\Bigr)\hat n = \omega_{p}\Bigl(\delta\bar{\rho}(x,t) -1\Bigr),\\
	& \text{where,}\quad  en_0\bar{\rho}(x,t) =\rho(x,t).
	\end{split}
\end{equation}
 Here, $\tau= \sqrt{\frac{\epsilon_0m}{n_0 q^2}}$ representing inverse of the plasma particle frequency $\omega_p$, which can be set to $1$ for even further simplification. It is a simple task to introduce multiple-ionised particles into these equations by replacing $\delta e$ with $\delta Z e$, $Z$ being the ionisation number. As shown in equation (\ref{vorticityrho}), vorticity associated with the plasma particle phase-space fluid flow is related to the local concentration of charge in a region of phase-space, defined by the spatial charge density $\rho(x)$. \\
 \indent The vorticity $\bm{\xi}$ associated with the phase-space, like any other ordinary fluid, describes the deformations in a region of the particle phase-space fluid. These deformations can be local sheering or rotational deformations, the latter of which describes a local spin of fluid elements about a common axis and is identified as a fluid vortex. These can also be defined as a local region of concentrated vorticity distinguished by closed or spiral streamlines \cite{Lugt1979}, or as described by \textcite{Tian2018a}, using the vortex parameter $g_{z}$ (or $g_{z\theta\rvert_{max}}$ in an invariant form) which can be defined for our system as --
\begin{subequations}\label{tianvalues}\allowdisplaybreaks
 		\begin{align}
 			\begin{split}
 		&g_z=\frac{\tau\partial \dot v_x}{\partial x}\frac{\partial \dot x}{\tau\partial v_x} = \frac{\delta e}{m}\frac{\partial E(x)}{\partial x}\frac{\partial v_x}{\partial v_x}\\&=\frac{\delta e}{m \epsilon_0}\rho(x)= \delta \omega_{p}^2 \bar{\rho}(x)
 		\end{split}\\
 	&g_{z\theta}\lvert_{\max}=\alpha^2 - \beta^2\\
 	\begin{split}=\frac{1}{4}\Biggl[\left(\frac{\tau\partial \dot v_x}{\tau\partial v_x}-\frac{\partial \dot x}{\partial x}\right)^2 + \left(\frac{\tau\partial \dot v_x}{\partial x} + \frac{\partial \dot x}{\tau\partial v_x}\right)^2 \\- \left(\frac{\tau\partial \dot v_x}{\partial x}-\frac{\partial \dot x}{\tau\partial v_x}\right)^2 \Biggr]\end{split}\\
 	&=\frac{1}{4}\Biggl[ \Bigl(\frac{\partial \dot x}{\partial x}+\frac{\partial \dot v_x}{\partial v_x}\Bigr)^2 - 4\frac{\partial \dot x}{\partial x}\frac{\partial \dot v_x}{\partial v_x} + 4\frac{\partial \dot v_x}{\partial x}\frac{\partial \dot x}{\partial v_x} \Biggr] \\
 	\begin{split}=\Biggl(\frac{\partial \dot v_x}{\partial x}\frac{\partial \dot x}{\partial v_x}-\frac{\partial \dot x}{\partial x}\frac{\partial \dot v_x}{\partial v_x} \Biggr)=\\\Biggl(\frac{\delta e}{m}\frac{\partial E(x)}{\partial x}\frac{\partial v_x}{\partial v_x}-\frac{\partial v_x}{\partial x}\frac{\delta e}{m}\frac{\partial E(x)}{\partial v_x} \Biggr)\\
 		=\frac{\delta e}{m}\frac{\partial E(x)}{\partial x} = \frac{\delta e}{m\epsilon_0}\rho(x) = \delta \omega_{p}^2\bar \rho(x) = g_z.\end{split}\\
 	\intertext{From ref. \cite{Tian2018a}, for a vortex region, }
 	 &g_{z\theta}\lvert_{\max}<0 \implies \delta \omega_{p}^2\bar \rho(x) <0 \implies \delta\bar \rho(x)<0.
 		\end{align}
 \end{subequations}
 \begin{figure}[!h]\centering
	\includegraphics[width=1.0\linewidth]{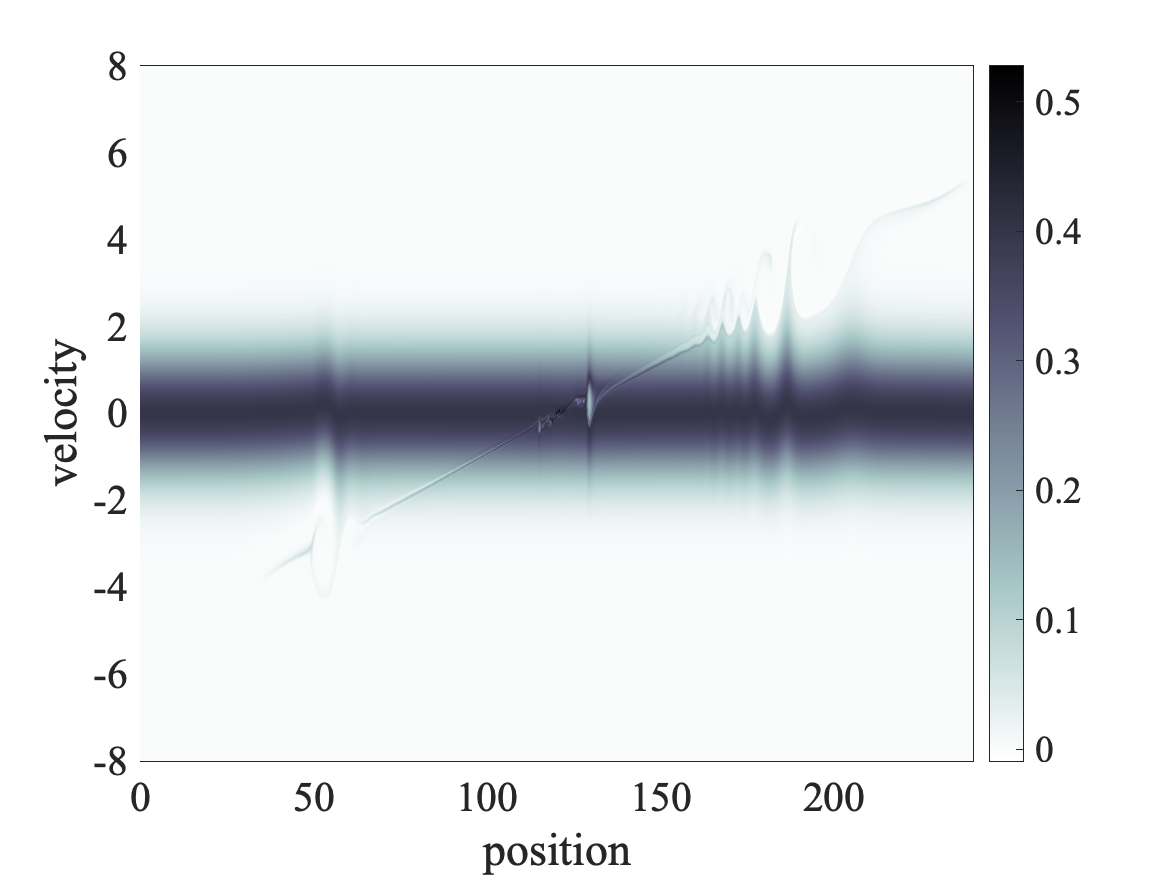}
	\caption{Phase-space portrait of the Q-machine Plasma showing electron hole formed using potential pulse with amplitude $A=3.0\frac{K_BT_e}{e}$, showing variation of phase-space density. Figure generated using kinetic simulation of the Q-machine experiment as performed by \textcite{Saeki1979}.}\label{figure2}
\end{figure}
 \\\indent The results described in equation (\ref{tianvalues}a,e,f) state the relation between the local  accumulation of charge in position-space and the rotational deformation caused by this accumulation in the phase-space fluid. It is well-known that phase-space electron and ion holes produce a well-defined local depression of particle charge in their regions. This would be in complete agreement with equation (\ref{tianvalues}f) since a vortical presence will be detected in any region of reduced particle charge, i.e.,
 \begin{equation}\label{vortexcondition}
 	g_{z\theta\rvert_{max}} =\begin{cases}
 	 \bar\rho(x)>0 &\text{for a -ve particle phase-space vortex}\\
 	 \bar\rho(x)<0 &\text{for a +ve particle phase-space vortex }
 	\end{cases}
 \end{equation}
 \\A key phenomena of this vortex region in the phase-space fluid model and its correspondence to the trapping condition must be included in the preceeding analysis -- the presence of fluid elements with both low and high velocities in the phase-space plane. Like any ordinary fluid, in the region of a vortex, fluid particles with low velocity produce complete rotational deformation, indicating the `trapping' of particles in the vortex. However, fluid elements with higher fluid velocity enter and exit the vortex region showing an arc-like trajectory.  Clearly, this trapping will be determined by the kinetic energy of the fluid element in comparison to the potential $\Phi$ developed due to the local accumulation of charge in the vortex, as shown in equation (\ref{tianvalues}a,e), i.e.,
 
 \begin{subequations}\label{trappingcondition}
 	\begin{align}
 	&\frac{1}{2}mv_x^2 + \delta e\Phi\geq 0 \implies \text{particles escape the vortex.}\\
 	&\frac{1}{2}mv_x^2 + \delta e\Phi<0 \implies \text{particle trapping.} 
 	\end{align}
 \end{subequations}
 
\begin{figure}[!ht]\allowdisplaybreaks
	\begin{subfigure}{2.9in}
		\includegraphics[width=1.0\linewidth]{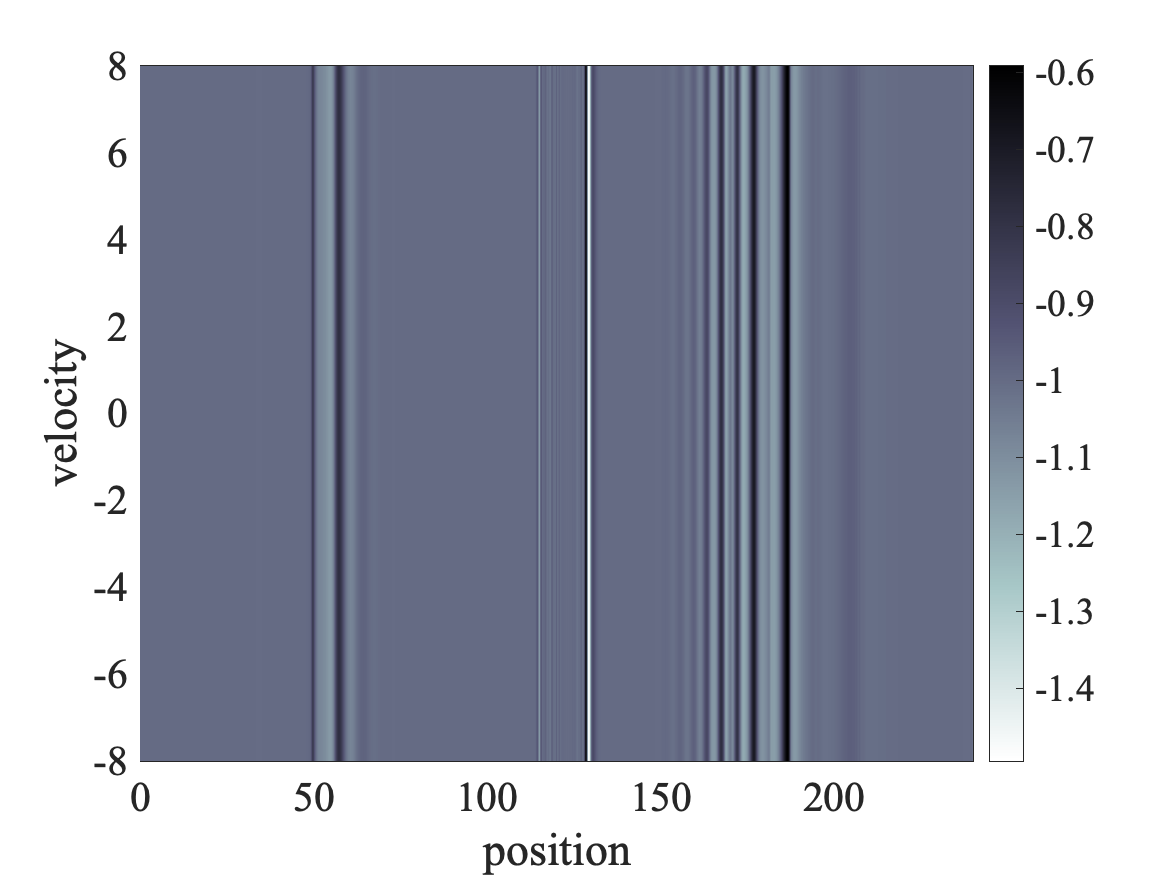}
		\caption{}
	\end{subfigure}
	\begin{subfigure}{2.9in}
		\includegraphics[width=1.0\linewidth]{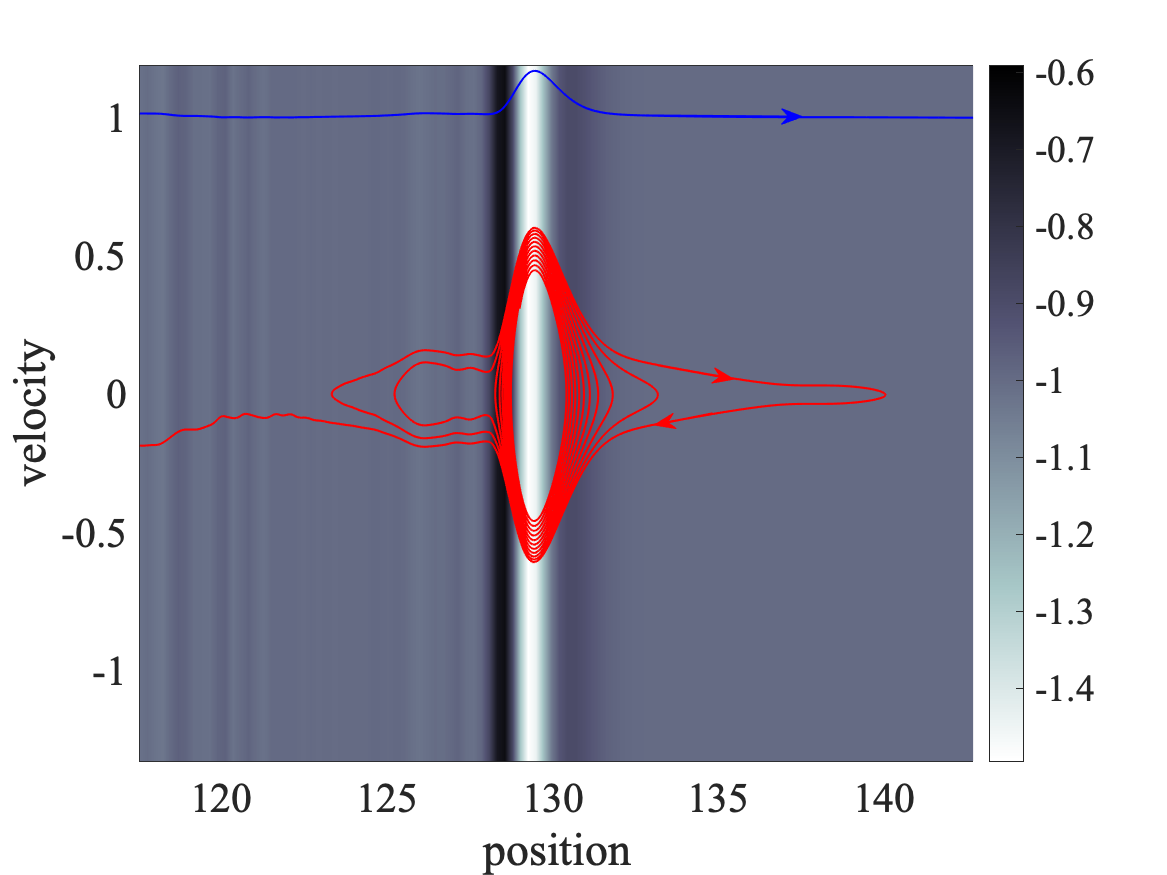}
		\caption{}
	\end{subfigure}
	\begin{subfigure}{2.9in}
		\includegraphics[width=1.0\linewidth]{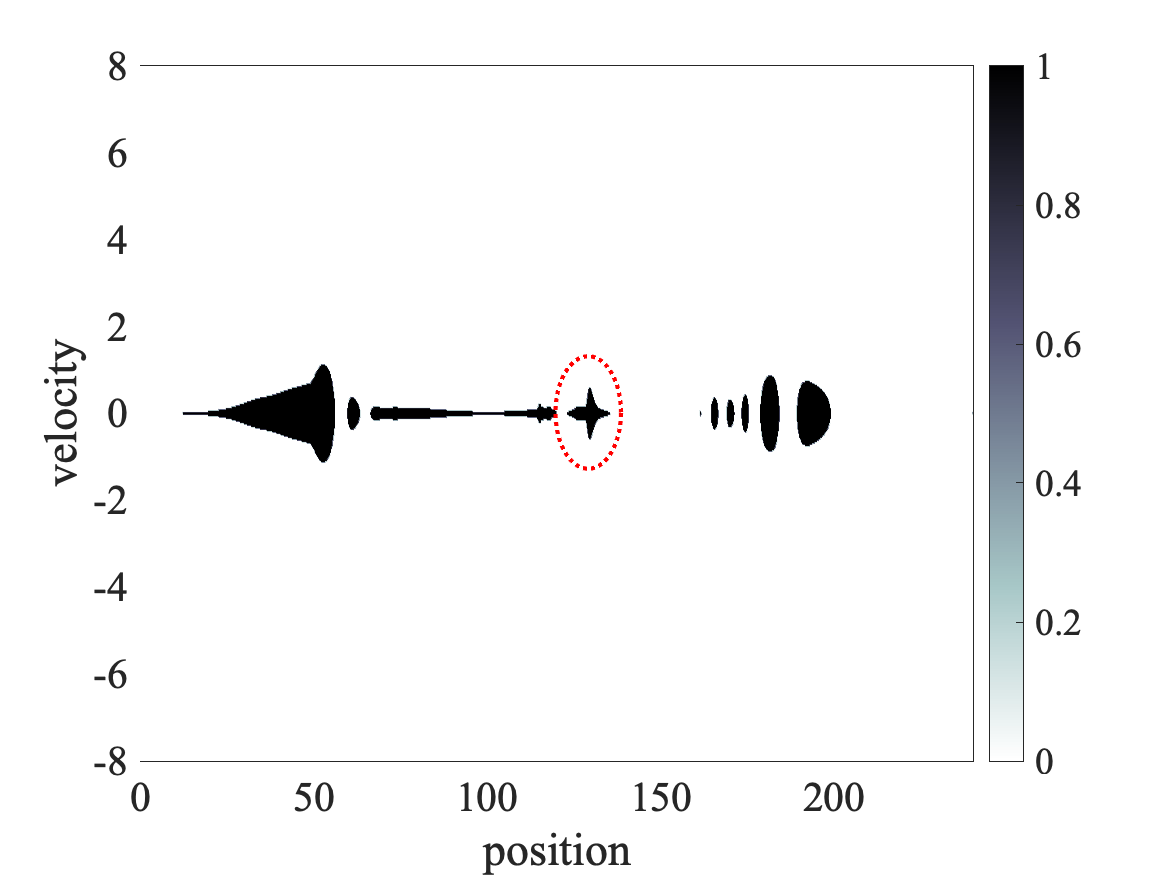}
		\caption{}
	\end{subfigure}
	\caption{(a) Vorticity of phase-space electron fluid, showing concentrated vorticity at the cite of the hole (see FIG. 3). (b) Streamline plots of electron phase-space fluid elements showing rotational streamline (red) in the region of the electron hole. (c) Map showing regions of rotational deformation (electron hole encircled in red) with $g_{z\theta\rvert_{max}}<0$. Note that other regions also identified as vortices due to formed turbulence produced by the numerical instability of the travelling compression and rarefaction waves.}\label{vortexidenti}
\end{figure}

\begin{figure}
    \centering
    \includegraphics[width=1\linewidth]{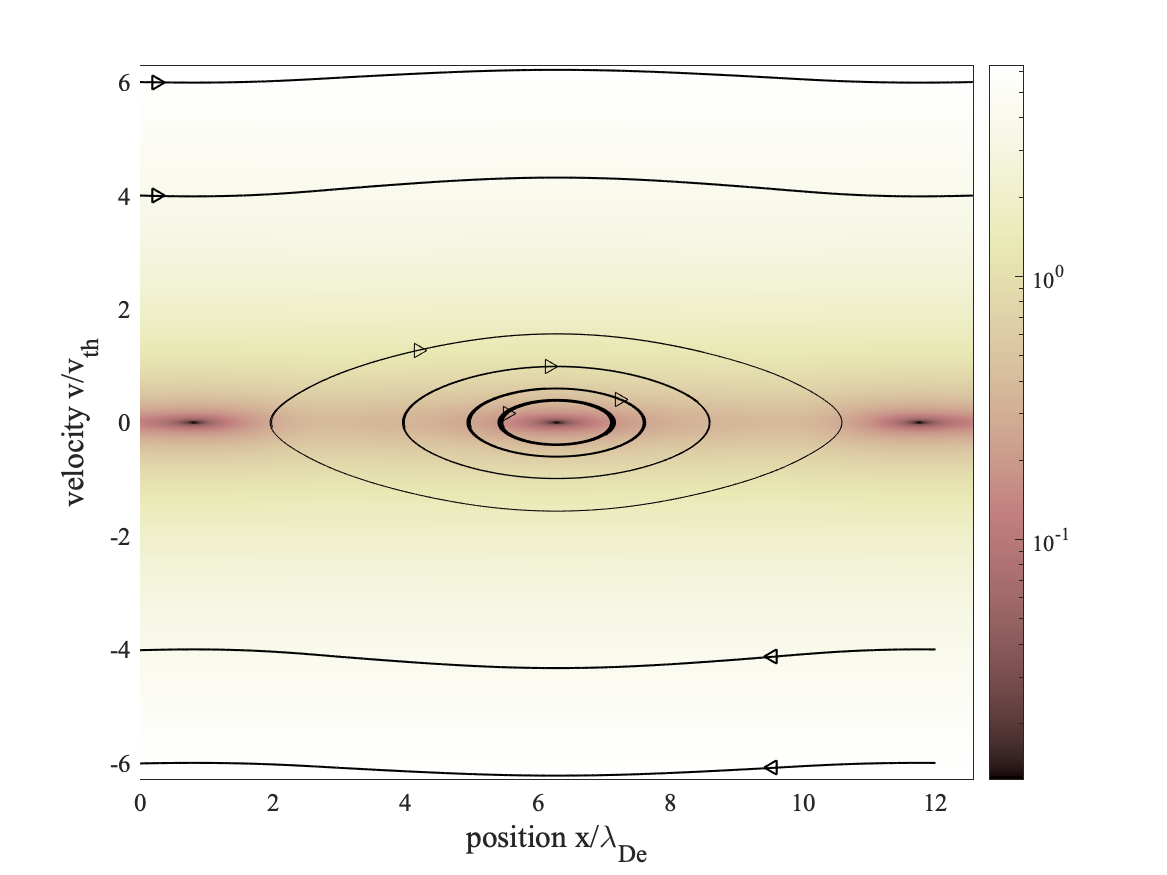}
    \caption{Velocity magnitude of an electron phase-space hole, in logarithmic scale. The nearly stationary core can be seen surrounded by fluid current rotating around it.}
    \label{fig:vmagofhole}
\end{figure}

 The conditions defined by equations (\ref{tianvalues}f), (\ref{vortexcondition}) and (\ref{trappingcondition}) can be used to identify vortices in the phase-space fluid, as shown in FIG. \ref{vortexidenti}. For this analysis, we numerically calculate the curl of the phase-space fluid velocity $\mathcal{V}$, as shown in equation (\ref{vorticity}a). Corresponding to FIG. \ref{figure2}, it can be seen that the vortcity in the hole region is concentrated and much larger than its surroundings, and streamlines of the phase-space fluid produce spiral and closed trajectories of the phase-space fluid element, which identifies that region as a vortex, according to \textcite{Lugt1979}. Furthermore, upon analysing the phase-space velocity magnitude $|\mathcal{V}|$ of the phase-space, it can be seen that the hole region is a vortex with a nearly stationary core with a fluid current rotating around it, with radially increasing velocity magnitude (see FIG. \ref{fig:vmagofhole}).\\
\indent We therefore conclude in this section that electron holes and ion holes are in-fact vortices in the phase-space of the collision-free plasma. Having found this analytical proof of their vortical nature, we can now approach the problem of electron phase-space holes using the fluid-like kinetic approach described in sec. II of this article. We do this in the succeeding section wherein we apply the phase-space-fluid model to study the structure of an electron phase-space hole.\\
\section{Analysis of the electron phase-space vortex structure in collision-less plasma phase-space}

In this section, we determine a steady-state equation for the phase-space density structure of this vortical region in the fluid phase-space. For the same, we use the fluid-momentum equation derived in equation (\ref{kineticeuler1}) along-with the isothermal pressure gradient introduced in equation (\ref{kineticpressuregrad}). In the region of a vortex, the temperature $T=T_e^{\text{tr.}}$, which is the temperature of the trapped electrons.\\
\begin{equation}\label{keuler}
		\eta m\left( \frac{\partial v_x}{\partial t} + v_x\cdot \nabla_x v_x\right)= \eta q E(x) - K_B T_e^{\text{tr.}}\nabla_x \eta
		\end{equation}
	We aim to find the steady state solution\footnote{Here, we have used the momentum equation along position-space (\ref{keuler}) which seems to be a more descriptive relation between the phase-space density function $f(x,v_x)$ and its coordinates, and not the Vlasov equation (\ref{continuityeqn}), which is the continuity equation of the phase-space fluid.} of this equation in the region of the electron vortex, by setting the material derivative of the phase-space fluid velocity (left side of equation (\ref{keuler})) equal to zero, as follows --
	\begin{subequations}\label{schamelderive1}\allowdisplaybreaks
		\begin{align}
		\eta m_e\frac{D( v_x)}{Dt}=0\Rightarrow\frac{q}{m_e\tau}fE(x) = \frac{1}{\tau}\nabla_x f\frac{K_B T_e^{\text{tr.}}}{m_e}\\
		\Rightarrow -fq\frac{\partial \phi(x)}{\partial x}=K_BT_e^{\text{tr.}}\frac{\partial f}{\partial x}\\
        - fq\frac{\partial \phi(x)}{\partial x}dx =K_BT_e^{\text{tr.}}\frac{\partial f}{\partial x}dx\Rightarrow  - fq d\phi =K_BT_e^{\text{tr.}}df\\
        \Rightarrow \int \frac{df}{f} =-\frac{q}{K_BT_e^{\text{tr.}}}\int \frac{\partial \phi}{\partial x}dx =-\frac{q}{K_BT_e^{\text{tr.}}}\int d\phi. 
	\end{align}
\end{subequations}
For the above integration, we set the limits from within the vortex region upto its boundary ( $0\leq v_x \leq \sqrt{2e\phi/m_e}$). For a constant Hamiltonian $H\leq 0$, change in potential can be expressed in terms of particle kinetic energy ($\mathcal{K}$) as follows --

\begin{subequations}\label{schamelderiv2}\allowdisplaybreaks
	\begin{align}
		&-qd\phi =  d\mathcal{K}\Rightarrow -q\Delta \phi = \Delta \mathcal{K} =q\phi - \frac{1}{2}mv_x^2 \\\intertext{Inserting in equation (\ref{schamelderive1}b) and normalising $\phi=\chi(K_B T_e q^{-1})$, $\mathcal{K}=\bar{v}_x^2 K_B T_e$, we get --}
	\begin{split}\ln f_{\text{bd.}}- \ln f_{\text{tr.}} = -\frac{1}{K_BT_e^{\text{tr}}}\left(\frac{1}{2}mv_x^2 - q\phi\right) \\
    = -\frac{K_BT_e}{K_B T_e^{\text{tr}}}\left(\bar v_x^2 - \chi\right)\end{split}\\
	&\Rightarrow \ln f_{\text{tr}} = -\left[(-\ln f_{\text{bd.}}) + \Bigl(-\frac{T_e}{T_e^{\text{tr}}} \Bigr)(\bar v_x^2 -\chi)\right].
    \intertext{Comparing with schamel-df equation for the trapped region\cite{Schamel1979}, we get --}
	\begin{split}\ln f_{\text{tr}} = -\left(M^2+ \beta(\bar v_x^2 - \chi)\right), \quad \text{where,}\quad  \\M^2 = -\ln f_{\text{bd.}},\qquad \beta = -\frac{T_e}{T_e^{\text{tr}}}.\end{split} \\
	& \Rightarrow \bar f_{\text{tr}} = \frac{1}{\sqrt\pi}\exp\left[-\bigl\{M^2+ \beta(\bar v_x^2 - \chi)\bigr\}\right].
	\end{align}
\end{subequations}

Using the same approach around the hole, it can be shown that the phase-space density remains a shifted Maxwellian in the un-trapped region --
\begin{equation}\label{freedf}
    \bar f_{\text{free}} = \frac{1}{\sqrt{\pi}}\exp\Bigl[ -\left(\bar M^2+\bar v_x^2 - \chi \right)\Bigr] , \qquad \bar v_x^2 \geq \chi.
\end{equation}

\indent The mathematics presented in equations (\ref{schamelderiv2}a-e, \ref{freedf}) reproduce the schamel-df\cite{Schamel1979} equations for an electron hole. These equations, which describe the phase-space density $f$ for both trapped and free electrons and were initially assumed B.G.K. solutions to the Vlasov-Poisson system, can be very-well derived upon treating the electron phase-space of collision-less plasmas as a fluid. \\

\begin{figure}[!h]
    \centering
    \includegraphics[width=1.0\linewidth]{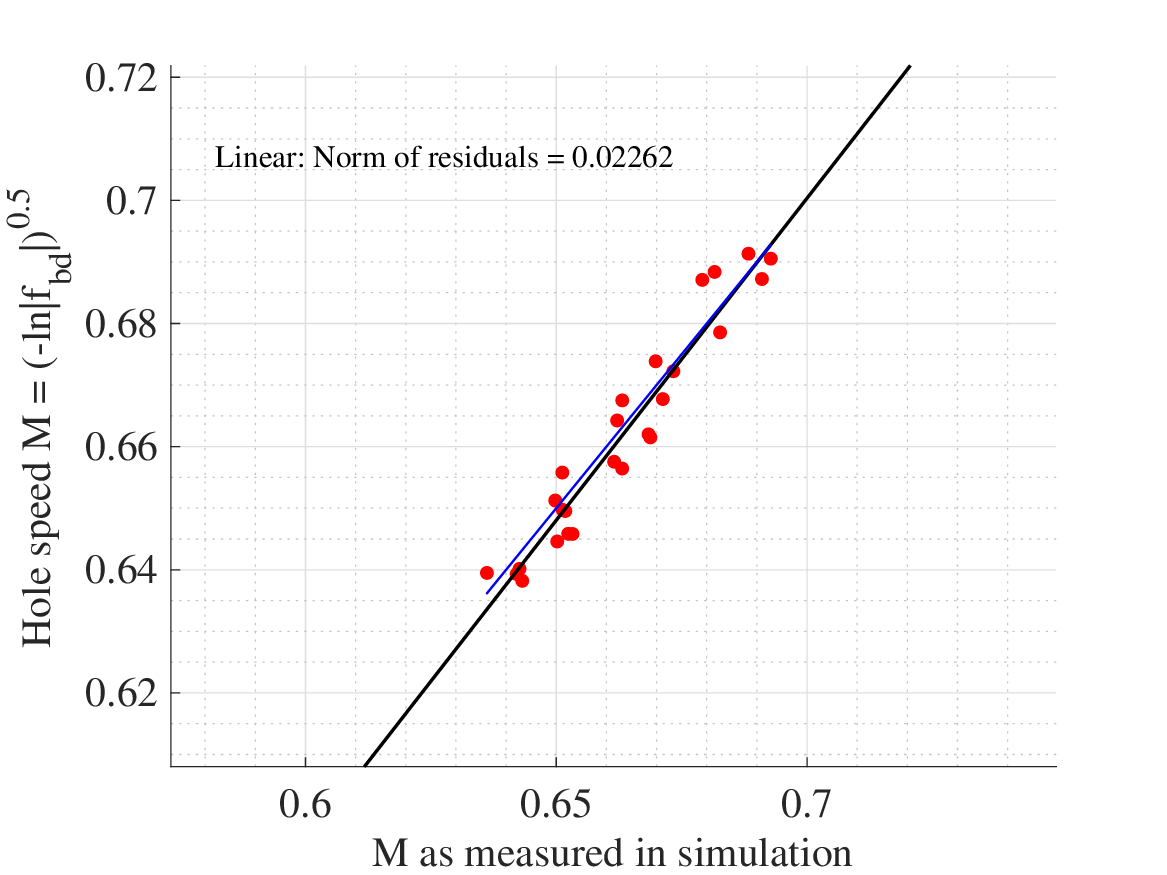}
    \caption{Hole speeds measured using kinetic simulation study of electron-hole formation in Q-machine plasma \cite{Saeki1979, Lynov1979, Lynov1980}. Relation between hole speeds recorded in simulation and the same calculated using relation  (\ref{M}) (red points), fitted curve (black line) and its comparison to the $x=y$ curve (blue line).}
    \label{fig:holespeed}
\end{figure}

The equations (\ref{schamelderiv2}a-e) also provide a relation between the normalised electron hole speed $M$ in the reference frame of the perturbation and its structure in phase-space (refer to FIG. \ref{fig:holespeed} ). The normalised electron hole speed $M$ is, as described in equation (\ref{schamelderiv2}d), equal to the square root of the phase-space density of the electron hole at its boundary, i.e.,
\begin{equation}\label{M}
	M=\sqrt{-\ln( f^{\text{tr.}}_{\text{bd.}})}=\sqrt{-\ln \Bigl[\sqrt{\pi}\bar{f}^{\text{tr.}}\bigl({x,v_x=\sqrt{{2e\phi}/{m_e}}}\bigr)\Bigr]}.
\end{equation}
\\
\indent Third, the definition of the particle trapping parameter $\beta$, as presented by Schamel in his works on particle trapping solitary B.G.K. waves \cite{Schamel1971, Schamel1972, Schamel1975, Schamel1979} can be extracted as a ratio of the plasma electron temperature to trapped electron temperature, as shown in equation (\ref{schamelderiv2}d). This is also in well-agreement to the initially assumed relation \cite{hutch2017}. Upon further analysis of the parabolic structure of this phase-space vortex, as clearly depicted from equation (\ref{schamelderiv2}c), a relation between this $\beta$ parameter and the depth of this vortex in the phase-space logarithmic surface can be deduced --
\begin{subequations}\label{betavalue}\allowdisplaybreaks
	\begin{align}
		\log_e\sqrt\pi\bar f_{tr} = -M^2 + \beta\chi -\beta \bar v_x^2\\
		\intertext{The minima of this parabolic surface is equal to $\ln f_{\text{min.}}^{\text{tr.}}$ at the centre of the hole, for which $\chi=\chi_0$ and $\bar v_x^2 =0$, with $\chi_0$ being the amplitude of the hole potential. Therefore,  }
		\beta = \frac{\ln f_{\text{min.}}^{\text{tr.}} + M^2}{\chi_0} = \frac{1}{\chi_0}\ln\left(\frac{f_{\text{min.}}^{\text{tr.}}}{f_{\text{bd.}}^{\text{tr.}}}\right) = \frac{\Gamma}{\chi_0},\quad \Gamma <0.
	\end{align}
\end{subequations}
The term $\Gamma = \ln{f_{\text{min.}}^{\text{tr.}}} - ln{f_{\text{bd.}}^{\text{tr.}}}$ in the above expression has been used to represent the (negative) depth of this phase-space vortex in the $\log_e f(x,v_x)$ surface. Thus,
\begin{equation}\label{depth-amplituderel}
    |\beta \chi_0| = |\Gamma|
\end{equation}
It is clear from this relation as to why the electron trapping parameter must be negative -- the value of $f_{\text{min.}}^{\text{tr.}}$ lies between $0$ and $f_{\text{bd.}}^{\text{tr.}}$, implying that $\Gamma \in (-\infty,0]$. Hence, $\beta$ must always be a non-positive value, its minima representing a hole with an empty core and maxima describing a flat surface electron hole. It is also clear from equation (\ref{depth-amplituderel}) that $\beta$ does not represent the hole depth in the phase-space plane ($-\Gamma$)\cite{Schamel2023PatternEquilibria}, but is actually the ratio of hole phase-space depth to the hole potential amplitude.\\

\begin{figure}
    \centering
    \includegraphics[width=1.0\linewidth]{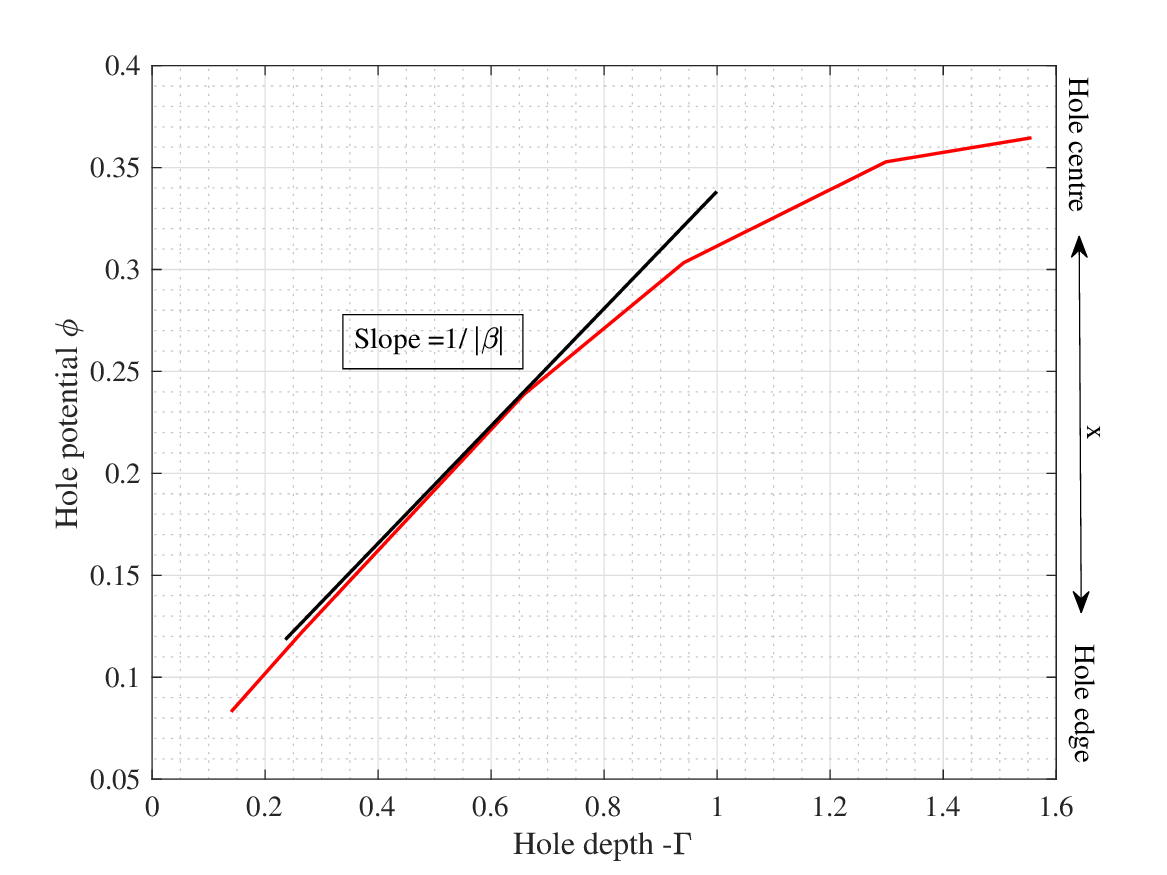}
    \caption{Variation of hole phase-space depth with hole potential across position-space (red curve) for a single hole. $1/|\beta|$ is the slope of the curve (approximated by the black line).  }
    \label{phigamma}
\end{figure}

Upon moving across an electron phase-space hole along position-space, one can observe that the hole phase-space depth ($-\Gamma$) must be directly proportional to the hole potential, keeping a constant $\beta$ for the hole. Thus, for a single hole, 

\begin{equation}
    \phi \propto -\Gamma.
\end{equation}
The same is shown in the numerical study of a solitary hole formed using the Q-machine plasma kinetic simulation\cite{Saeki1979} (refer to FIG. \ref{phigamma}).\\

\section{Discussions and Conclusion}
In this article, a fluid outlook of the phase-space has been explored. This approach, as has been shown, analytically confirms the vortical nature of the phase-space holes, as was initially stated by many authors. This implies the validity not just of their vortical nature, but also of the applicability of a fluid-like treatment in the study of these phase-space vortices using a different mathematical framework -- continuity (Vlasov) as well as momentum equation system of the phase-space. It confirms that when the phase-space is observed in analogy to a two-dimensional fluid surface, a velocity field and a vorticity field associated with the flow-like dynamics of the phase-space element can be defined. Various contortions can then be analyzed using these fields and analogous to the analytical techniques of two-dimensional fluid dynamics, the phase-space characteristics of such kinetic structures can be studied.\\
\indent The study of electron holes using this approach reproduces the schamel-df equations \cite{Schamel1979} which depict the phase-space structure of both free and trapped particle densities in the electron hole region. Further, it derives the definition of the electron trapping parameter $\beta$ as it was originally described \cite{Schamel1979, schamel1986electron, hutch2017}. It also provides an analytical relation between electron hole speed $M$, potential amplitude $\chi_0$, the trapping parameter $\beta$ and the hole depth in its phase-space $(-\Gamma)$. This is a measurable definition of the electron trapping parameter.\\
\indent Hence, this approach provides a precursor to the B.G.K. differential approach -- allowing one to not assume by analytically derive a well-defined particle distribution function for the study of the phase-space vortices using the pseudo-potential method \cite{Schamel1979, schamel1986electron, Schamel2023PatternEquilibria}.\\
\indent It is important to note that the stochastic nature of the studied kinetic phenomena and the abundance in the trapping mechanisms can be accounted for in this model upon further exploration and modification of the phase-space continuity-momentum equation system, as well as modification of intrinsic plasma conditions. We, however, leave such explorations for further development of this model and study.
\section*{Appendix}
\renewcommand{\theequation}{A.\arabic{equation}}
\setcounter{equation}{0}
The self-consistency of a fluid requires solving a Laplace equation of the stream function $\mathcal{H}(x,y)$, such that\cite{schamel2012cnoidal} --
    \begin{equation}\label{streamer}
        \mathcal{V}(x,y) = \nabla \mathcal{H} \times \hat{n},
    \end{equation}
    where, $\hat{n}$ is a direction normal to the two-dimensional fluid. The same, in the case of the fluid-analogue of the particle phase-space, can be described as a modified Hamiltonian $H$ of the particles --
    \begin{equation}\label{streamfunc}
        \mathcal{H} = \frac{\tau}{m}H = \frac{\tau}{m}\left(\frac{1}{2}mv_x^2 + q\phi(x)\right).
    \end{equation}
    Here, $\phi(x)$ is the potential field experienced by the particle of charge $q$. Taking $\hat n$ to be normal to the phase-space plane and using $(\hat x, \hat v_x, \hat n)$ as a complete orthogonal set such that $\hat x\times \hat v_x = \hat n$, and inserting equation (\ref{streamfunc}) in equation (\ref{streamer}), we see that --
    \begin{subequations}\allowdisplaybreaks
        \begin{align}
            \nabla \mathcal{H} &= \frac{\tau}{m}\left( q\frac{\partial \phi}{\partial x}\hat{x} + \frac{m}{2\tau}\frac{\partial v_x^2}{\partial v_x}\hat{v}_x\right)\\
            &=\frac{q\tau}{m}\frac{\partial \phi}{\partial x}\hat{x} + v_x\hat{v}_x.\\
            \nabla \mathcal{H}\times\hat{n} &=\frac{q\tau}{m}\frac{\partial \phi}{\partial x}(\hat{x}\times\hat{n}) + v_x(\hat{v}_x\times\hat{n}).\intertext{From $\hat x\times\hat v_x = \hat n$, we get--}
            \hat x\times\hat n &= -\hat v_x,\quad \hat v_x \times\hat n=\hat x.\\
            \therefore\nabla \mathcal{H}\times\hat{n} &=\frac{q\tau}{m}\left(-\frac{\partial \phi}{\partial x}\right)\hat{v}_x + v_x\hat{x} = \mathcal{V}.
        \end{align}
    \end{subequations}
    Therefore, upon using the relation between $\hat{n}$, $\hat{x}$ and $\hat{v}_x$ defined in equation (\ref{vorticity}b), equation (\ref{streamer}) is satisfied. The stream function (\ref{streamfunc}) can then be acted upon by the Laplacian operator in phase-space to reproduce the vorticity expression defined in equation (\ref{vorticity}) --
    \begin{subequations}
        \begin{align}
            \nabla \cdot \nabla\mathcal{H} &=\frac{q\tau}{m}\frac{\partial^2 \phi}{\partial x^2} + \frac{1}{\tau}\frac{\partial v_x}{\partial v_x}\\
            &=-\left(\frac{\tau}{m}\frac{\partial \bm{F}(x)}{\partial x} - \frac{1}{\tau}\right) = - \bm{\xi}(x),\\
            \text{where,}&\quad \bm{F}(x)=-q\frac{\partial\phi}{\partial x}\hat{x}.
        \end{align}
    \end{subequations}
    This shows the self-consistency of the fluid-analogue of the particle phase-space.
\bibliography{authors.bib}
\bibliographystyle{apalike}

\end{document}